# Coupled gyration modes and propagation in one-dimensional periodic skyrmion arrays as reliable information carrier

Junhoe Kim, Jaehak Yang, Young-Jun Cho, Bosung Kim, and Sang-Koog Kim [a]

*National Creative Research Initiative Center for Spin Dynamics and Spin-Wave Devices, Nanospinics Laboratory, Research Institute of Advanced Materials, Department of Materials Science and Engineering, Seoul National University, Seoul 151-744, Republic of Korea*

[a] Correspondence and requests for materials should be addressed to S.-K.K. (sangkoog@snu.ac.kr).

We report on a micromagnetic numerical simulation study of dynamic couplings between neighboring skyrmions in narrow-width nanostrips. We explored the coupled gyration modes and their characteristic dispersions in terms of the interdistance between the neighboring skyrmions. The application of perpendicular magnetic fields, importantly, allows for control/modification of the dispersion of the coupled modes. Coupled skyrmion gyration modes provide a new type of efficient, reliable, fast, low-power-consumption information-signal propagation in narrow-width straight and curved nanostrips, as driven predominantly by the exchange interaction between individual nano-scale skyrmions.

Topologically stable magnetic skyrmions[1,2] have been observed in both bulk magnetic materials of non-centrosymmetry[3,4] and magnetic thin films of broken inversion symmetry at hetero-interfaces of large spin-orbit coupling.[5-7] It is well known that the Dzyaloshinskii-Moriya interaction (DMI)[8,9] plays a crucial role in stabilizing the unique spin textures of skyrmions in bulk and thin-film materials.[3-7] Skyrmions' characteristic features, including nano-scale size, topological stability, and ultra-low threshold current density necessary for their motions, make them promising potential candidates for information-storage and -processing device applications.[6,10-15] Not surprisingly then, recent observations of room-temperature magnetic skyrmions have attracted further interest for the fundamental and technological implications.[16-18]

Additionally to such exotic spin textures, several fundamental dynamic modes of skyrmion crystals have been found theoretically by M. Mochizuki[19] and also experimentally by Y. Onose et al.[20] Both revealed the existence of skyrmion-core gyration modes of either the clockwise (CW) or counter-clockwise (CCW) rotation sense, as excited by in-plane ac magnetic fields[19,20], as well as another breathing mode[19,20] excited by out-of-plane ac magnetic fields. Such internal dynamic modes also have been found in single skyrmions in infinite films[21] or in confined geometries.[22] Further, collective excitations in 1D chains of single-skyrmion nanodisks[23] along with spin-wave propagations and their dispersion in 1D periodic skyrmion lattice nanostrips[24] have been investigated as well.

Whereas the above studies focused on the fundamental dynamic modes of single skyrmions, 1D skyrmion arrays and 2D/3D skyrmion crystals or coupled skyrmion gyration modes in 1D periodic skyrmion lattices in narrow-width nanostrips have yet to receive much attention in terms of the implications for reliable information carriage in straight and curved nanostrips without skyrmion motion. In the current work therefore, we not only explored the gyration modes of two, five and more coupled skyrmions in continuous thin-film narrow-width nanostrips but also elucidated the underlying physics of those coupled modes. Their characteristic dispersions as well as manipulations by external perpendicular fields were investigated in the aspect of their potential applications as reliable information carriers.

## Results

**Intrinsic dynamic modes of single skyrmion in nano-square dot.** In the simulation, first, to determine the intrinsic dynamic modes of an isolated skyrmion, we employed, as shown in Fig. 1(a), a single skyrmion of downward core orientation in a 40 nm-wide square dot of thickness of $h$ = 1 nm. Sinc-function fields of $H = H_0 \sin[\omega_H(t-t_0)]/[\omega_H(t-t_0)]$ with $H_0$ = 10 Oe, $\omega_H = 2\pi \times 50\,\text{GHz}$, and $t_0$ = 1 ns were applied along the $y$ (in-plane) and $z$ (out-of-plane) axis during a $t$ = 100 ns time period. Figure 1(b) shows the spectra obtained from the fast Fourier transform (FFT) of the temporal oscillations of the $y$ component of the skyrmion-core position[25,26] under an in-plane ac magnetic field (upper panel), as well as the temporal oscillations of the $m_z = M_z/M_s$ component averaged over the entire area of the square dot under an out-of-plane ac field (bottom). As can be seen, two distinct peaks were found at 1.28 and 22.3 GHz, which correspond to the eigenfrequencies of the CW gyrotropic and breathing modes, respectively. These two single-skyrmion modes are consistent with the CW gyrotropic and breathing modes in a confined disk, as found in earlier work.[22,23] The gyration mode of an isolated skyrmion also is similar to the well-known gyration mode of a single vortex in soft nanodots.

**Coupled modes of two skyrmions in rectangular dot.** In previous work, it was reported that the gyrations of single vortices can be coupled to form collective modes in magnetically coupled vortices or vortex-antivortex systems. Such coupled modes, significantly from the technological point of view, can be used as information carriers, as has already been reported [27-29]. Therefore, in the present study, we also took into account dynamic coupling between two different skyrmions, as shown in Fig. 2(a). The two skyrmions were coupled via dominantly short-range strong exchange interaction, because each of the two core motions affects the other, whereas vortices between physically separated disks are coupled via a relatively weak dipolar interaction, without direct physical contact between vortices.

To excite all of the excitable modes in the two skyrmions, we applied a static field of 700 Oe strength in the $+x$ direction locally to the first skyrmion (noted as Sky1) region in order to displace its initial core center ~1 nm in the $-x$ direction. After turning off the local field, we traced the trajectories of both skyrmion centers under free relaxation. The simulation results noted hereafter were obtained up to 200 ns after the field was turned off. The trajectories of the individual core-position vectors $\mathbf{X} = (X, Y)$ are plotted in Fig. 2(b) along with their power spectra in the frequency domain as obtained from the FFTs of the $y$ component of the individual core oscillations, as shown in Fig. 2(c). For both skyrmions, two peaks, lower and higher, denoted $\omega_l$ and $\omega_h$ and corresponding to $2\pi \times 0.88$ and $2\pi \times 1.48$ GHz, respectively, were observed. From the inverse FFTs of each peak (mode) for each skyrmion, we obtained the spatial correlation between the two core motions for the lower (top) and higher mode (bottom), as shown in Fig. 2 (d). For the $\omega_l$ mode, the two cores move in phase, while for the $\omega_h$ mode, the two cores move in antiphase. For both downward cores, the core motion is in the CW rotation sense, the same as that of the downward vortex core. It was reported in our previous works[30-32] that two normal modes in two dipolar-coupled vortices appear due to the breaking of the radial symmetry of the potential wells of decoupled isolated disks by the dipolar interaction between the two vortex cores displaced from their respective center positions. Analogously, the mode splitting and coupled skyrmion-core gyration in the skyrmion pair are explained by the symmetry breaking of the potential energy of isolated skyrmions. However, the exchange interaction is dominant in such skyrmion dynamic coupling in perpendicularly magnetized thin films. Also, the elliptical trajectories of both skyrmion cores were observed: for the $\omega_l$ mode the major axis of the elliptical shape is along the *x* (bonding) axis, while for the $\omega_h$ mode, it is along the *y* axis.

**Coupled modes in five-skyrmion array in nanostrip.** On the basis of the above results, we extended our simulations to a five-skyrmion chain in a rectangular-shape nanostrip [Fig. 3(a)]. We excited all of the fundamental modes in the given system by applying a static magnetic field of $H_x$ = 700 Oe only to the first skyrmion (left end), in the same way as earlier. After the field was turned off, the signal of the gyrations excited from the first skyrmion core propagates to the next skyrmion and reaches the 5$^{th}$ skyrmion at the other end under free relaxation. Because all of the five cores' motions were coupled, the resultant gyration trajectories of the individual cores were complex, as shown in Fig. 3(b). To better understand the observed complex core trajectories (as superposed by the five different modes), we also plotted, in Fig. 3(c), the frequency spectra for the individual core motions. The five distinct peaks corresponding to the five internal modes are denoted $\omega_i$ ($i$ = 1, 2, 3, 4, 5). For each skyrmion core, the five peaks in the frequency domain are at $\omega/2\pi$ = 0.40, 0.79, 1.13, 1.19, and 1.38 GHz. For each skyrmion core too, the five peaks are located at the same position, though the FFT powers, from the first skyrmion through the fifth, contrast. For the first and fifth skyrmions, all of the five peaks appear. By contrast, for the 2nd and 4$^{th}$ skyrmions, the $\omega_3$ peak disappears, while for the 3rd skyrmion, the $\omega_2$ and $\omega_4$ peaks disappear. These spectra together represent the characteristic intrinsic modes of collective motion for the given whole system.

To interpret the complex core motions for each of the five different $\omega_i$ modes, we formulated inverse FFTs of the individual cores' positions. The resultant trajectories of the orbiting cores and the corresponding *y*-component profiles for each mode are illustrated in Fig. 4. On the basis of a fixed boundary condition[27], the wave vector of the allowed modes is expressed as $k = m \cdot \pi / [(N+1)d_{int}]$, with *N* the number of skyrmions in a nanostrip, $d_{int}$ the interdistance between the neighboring skyrmions, and *m* a positive integer subject to the constraint $m \leq N$. Thus, the discrete five modes' *k* values of collective skyrmion-core gyration are given as $k_m = m\pi / 6d_{int}$, where *m*=1, 2, 3, 4, 5, indicating each mode. The collective motions of the individual five skyrmion cores show unique standing-wave forms of different wavelengths, $2\pi / k_m$. The orbiting radii of the five cores are symmetric with respect to the center of the whole system (i.e., at the 3rd skyrmion) and are also completely pinned at the imaginary points 7 nm from both ends, as reported in Refs. 27-29. For the lowest mode, $\omega_1$, all of the cores gyrate in-phase. As the wavelength of those standing waves decreases, the relative phases between the nearest-neighboring cores are increased, and thus, the standing-wave nodes appear. The nodes can be observed only under the $\sin(m\pi / [(N+1)d_{int}])\hat{\mathbf{k}} \cdot nd_{int}\hat{\mathbf{x}}) = 0$ condition, according to which the nodes of the standing waves are located inside the $n^{th}$ disk.

**Dispersion in 1D skyrmion lattice in nanostrip.** In order to examine the dispersions of such collective motions in 1D coupled finite-number skyrmions, we conducted further simulations of a model of longer 1D chains consisting of *N* = 25 skyrmions in a nanostrip of the following dimensions: length *l* = 800 nm, width *w* = 40 nm, thickness *h* = 1 nm (see Fig. 5(a)). The first skyrmion core was shifted and then allowed to relax in the same way as earlier. Figure 5(b) provides a contour plot of the *y*-axis displacements of the individual cores from their center positions with respect to position and time. The excited gyration of the first skyrmion core propagates well along the whole nanostrip, as evidenced by the 1st wave-packet propagation through the entire skyrmion chain. The speed of the gyration-signal propagation indicated by the black line is ~ 135 m/s, which value is generally two times faster than that for a 1D vortex-state disk array.[28]

From the FFTs of the temporal oscillations of the *y* components of the individual cores' position vectors, we also obtained dispersion relations in the reduced zone scheme, as shown in Fig 6. As is evident, the dispersion curves are asymmetric with respect to $k = 0$; the intensities of the modes with positive group velocities are higher than those with negative group velocities, because the gyration signal was excited from the left end and then was allowed to propagate in the +*x* direction; the overall shape of dispersion was concave up (that is, the frequency was lowest at $k = 0$ and highest at $k = k_{BZ} = \pi/d_{int}$); at $k = 0$, all of the cores move together coherently, while at $k = k_{BZ}$, they act as the nodes of the standing wave. Such collective dynamic motions are determined predominantly by a strong exchange interaction according to the relative positions of the nearest-neighboring skyrmion cores, as shown in Supplementary Materials Fig. 1.

**Dependences of the dispersion of 1D skyrmion chains on $d_{int}$ and $H_z$.** Next, in order to examine the dependence of the band structure of 1D skyrmion lattices on $d_{int}$, we varied the skyrmion number *N* from 21 to 29 for the given nanostrip dimensions, as shown in Fig 6(a). In the results, stable skyrmion lattices were obtained from $d_{int} = 38$ to 27 nm, according to *N*. The contrasting band structures for the different $d_{int}$ values are shown in Fig. 6(a). It can be seen that as $d_{int}$ decreases, the band width $\Delta\omega$ and the angular frequency $\omega_{BZ}$ at $k = k_{BZ}$ increase (see

Fig. 7(a), left). This increase in $\Delta\omega$ and $\omega_{BZ}$ with decreasing $d_{int}$ can be explained by the variation of the magnetic energy with $d_{int}$. Whereas the dipolar, anisotropy and DMI energies are not much influenced by $d_{int}$, the exchange energy increases remarkably with decreasing $d_{int}$, as shown in Supplementary Fig. 1. Moreover, the temporal variation of the exchange energy density is two times larger than that of the magnetostatic energy during the skyrmion cores' motion, which indicates that the exchange interaction plays the dominant role in the coupled gyration modes of skyrmions. This means, further, that the $d_{int}$-dependent band structure is correlated with magnetic interactions, specifically exchange interaction.

Here, in additional simulations applying perpendicular magnetic fields of different strengths $H_z$ = +2, +1, -1 and -2 kOe, we also demonstrated an external control of the band structure of skyrmion lattices. The number of skyrmions in the nanostrip was set as $N$ = 25 (i.e., $d_{int}$ = 32 nm). Figure 6(b) compares the dispersion curves for indicated different $H_z$ values, showing clearly that the bandwidths of the resultant band structures decrease with increasing $H_z$. As seen in Fig. 7(a)'s $\omega_{BZ}$ versus $H_z$ plot, the $\omega_{BZ}$ also decreases linearly with increasing $H_z$. The application of the $H_z$ field modifies the magnetization profiles of each of the skyrmions' cores, as shown in Supplementary Materials Fig. 2. That is, the skyrmion shrinks with the positive external magnetic fields and expands with the negative external magnetic fields. Accordingly, the eigenfreqeuncies $\omega_0$ of the skyrmion's CW rotation vary with $H_z$.[19,33,34] Based on further micromagnetic simulations, we also confirmed that the CW mode's eigenfrequency of a single skyrmion changes linearly with $H_z$ (see Supplementary. Fig. 3) Therefore, the linear dependences of $\omega_{BZ}$ on $H_z$ are mostly associated with the variation of the $\omega_0$ of the isolated skyrmions. In fact, the $\omega_{BZ}$ value corresponds to the $\omega_0$ of isolated skyrmions, as shown in Supplementary Fig. 3.

## Discussion

From a technological point of view, such gyration-signal propagation in a 1D skyrmion array can be used as a reliable information carrier. Figure 7(b) shows the gyration-signal propagation speed estimated from the displacement of individual cores from their center positions for different values of $d_{int}$ (left column) and $H_z$ (right). The resultant propagation speeds generally follow the dependence of $\omega_{BZ}$ on $d_{int}$ (left column) and $H_z$ (right). The above results are promising for potential signal-processing applications, because of the following advantages. First, 1D periodic skyrmion lattices at room temperature can be simply obtained in nanoscale-width nanostrips. Second, the creation or annihilation of single skyrmions as well as the periodicity of skyrmion lattices can be manipulated using simple methods such as entails the application of a magnetic probe tip or by control of either the frequency of the magnetic field or the spin-polarized current pulse in nanotracks of different width.[7,35] Third, the propagation speed is controllable with an externally applied perpendicular field. Forth, the most important benefit of using the skyrmion gyration modes as information carriers is the fact that skyrmion gyrations can be excited with extremely low power consumption (e.g., an ac resonant field of a few tens of Oe or less,[19] or currents[11] on the order of ~$10^6 A/m^2$). Fifth and finally, such a fast gyration signal propagates even in curved nanostrips (e.g., L or round corners with no edge modifications). We performed additional simulations of a 1D skyrmion array in an L-shaped nanostrip, as shown in Supplementary Fig. S4. It was found that the gyration signal excited from one end propagates well to the other end through the L-shaped nanostrip. In previous work,[13,36,37] it has been reported that in such curved structure, skyrmion annihilation occurs at the corner edge when the skyrmion motion is driven by currents or spin-waves. Therefore, the method proposed in this paper, notably, is very advantageous for reliable logic operation in complex structures.

In summary, we explored the gyration modes of coupled skyrmions and their dispersions in 1D skyrmion lattices. The modes and their characteristic dispersion relations were examined for different skyrmion interdistances and perpendicular magnetic fields externally applied to the nanostrips. Additionally, the controllability of the dispersion curves and skyrmion gyration propagation were demonstrated. The strong exchange coupling between neighboring skyrmions leads to the propagation of skyrmion gyrations as fast as ~135 m/s, which value, significantly, are controllable by applied perpendicular fields. This work provides not only a fundamental understanding of the dynamics of coupled skyrmions but also a new type of skyrmion magnonic crystal applicable to future information processing devices.

## Methods

**Micromagnetic simulations.** To numerically calculate the dynamic motion of local magnetizations in nanostrips of skyrmion lattices, we employed the Mumax3 code[38] that utilizes the Landau-Lifshitz-Gilbert (LLG) equation,[39,40] $\partial \mathbf{M}/dt = -\gamma \mathbf{M} \times \mathbf{H}_{\text{eff}} + (\alpha/M_s)\mathbf{M} \times \partial \mathbf{M}/\partial t$, where $\gamma$ is the gyromagnetic ratio, $\alpha$ the damping constant, $\mathbf{H}_{\text{eff}}$ the effective field, and $M_s$ the saturation magnetization. For a model system of Co thin films interfaced with heavy metal (Pt) and of perpendicular magnetic anisotropy, we used the following material parameters: $M_s$ = 580 kA/m, exchange stiffness $A_{ex}$ = 15 pJ/m, perpendicular anisotropy constant $K_u$ = 0.8 MJ/m$^3$, DMI constant $D$ = 3.0 mJ/m$^2$.[14] The unit cell was $1.0 \times 1.0 \times 1.0$ nm$^3$ in its dimensions. A damping constant of $\alpha$ = 0.3 was used to obtain the initial ground states of the skyrmions, though a much smaller value, $\alpha$ = 0.0001, was used in order to examine their dynamic motion and achieve, thus, a better spectral resolution.

## Acknowledgments

This research was supported by the Basic Science Research Program through the National Research Foundation of Korea (NRF) funded by the Ministry of Science, ICT & Future Planning (NRF-2015R1A2A1A10056286).

## Author Contributions Statement

S.-K.K. and J.K. conceived the main idea and planned the micromagnetic simulations. J.K. performed the micromagnetic simulations. J.K., J.Y., Y.-J.C., B.K., and S.-K.K. analyzed the data. S.-K.K. led the work and wrote the manuscript with J.K. The other co-authors commented the manuscript.

## Additional information

The authors declare no competing financial interests.

# Figure Legends

**Figure 1 | Intrinsic modes of a single skyrmion in a nano-square dot.** (a) Single skyrmion in perpendicularly magnetized nano-square dot of indicated dimensions. The colors correspond to the out-of-plane magnetization components $m_z = M_z/M_s$. (b) FFT power spectra in frequency domain, as obtained from fast Fourier transform (FFT) of temporal oscillations of $y$ position of skyrmion-core motion excited by in-plane sinc field (top) and oscillation of $m_z = M_z/M_s$ averaged over entire nano-square dot excited by out-of-plane sinc field (see the text for the form of sinc field used).

**Figure 2 | Normal-mode representation of two coupled skyrmions in a rectangular dot.** (a) Two coupled skyrmions in rectangular dot of indicated dimensions. (b) Trajectories of skyrmion-core motions in period of $t = 0$- 100 ns. (c) FFT power spectra obtained from FFTs of $y$ components of two core-position vectors from their own center positions. (d) Trajectories of two core motions in one cycle period ($2\pi/\omega$) of the gyration for each mode. The blue dots on the trajectory curves represent the positions of the individual cores.

**Figure 3 | Normal-mode representation of five coupled skyrmions in a nanostrip.** (a) Five-skyrmion chain in perpendicularly magnetized nanostrip. (b) Trajectories of gyration motions of individual cores of five skyrmions and (c) their FFT spectra. The five distinct peaks are denoted as the individual mode $\omega_i$ peaks ($i = 1, 2, 3, 4,$ and $5$).

**Figure 4 | Spatial correlation of the five cores' gyration motions for each mode.** The red dots indicate the core positions of the individual skyrmions. The trajectories are magnified for clear comparison, and the magnifications differ for the individual modes. The corresponding profile of the *y* components is indicated by the solid line in the right panel.

**Figure 5 | Propagation of excited skyrmion gyrations through 1D skyrmion array.** (a) 1D skyrmion array in nanostrip of indicated dimensions comprising 25 skyrmions. (b) Contour plot of *y* components of individual cores' displacements with respect to time and distance in whole chain. The zero position is defined as the first skyrmion-core position in the initial state. The color bar indicates the *y* components of the individual cores' displacements from their own initial positions.

**Figure 6 | Dependences of the dispersion of 1D skyrmion arrays on interdistance $d_{int}$ and perpendicular magnetic field $H_z$.** (a) Dispersion curves of 1D skyrmion chains for different interdistances, e.g., $d_{int}$ = 27, 29, 32, 34, 38 nm. (b) Dispersion curves of 1D skyrmion chains of 25 skyrmions ($d_{int}$=32 nm) for different perpendicular fields, e.g., $H_z$ = -2, -1, 0, 1, 2 kOe.

**Figure 7 | Angular frequency $\omega_{BZ}$ at $k=k_{BZ}$, and propagation speed of coupled gyrations versus $d_{int}$ (left panel) and $H_z$ (right)**

**Figure1**

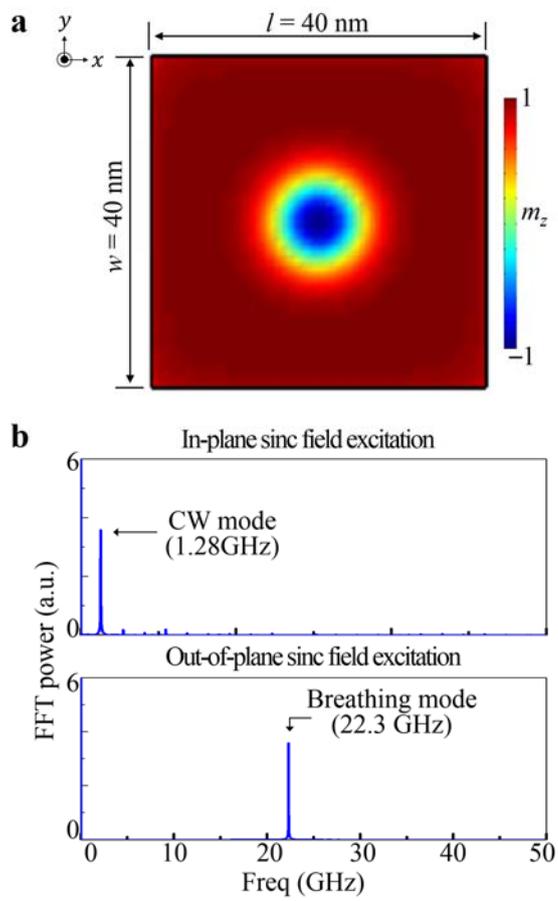

**Figure2**

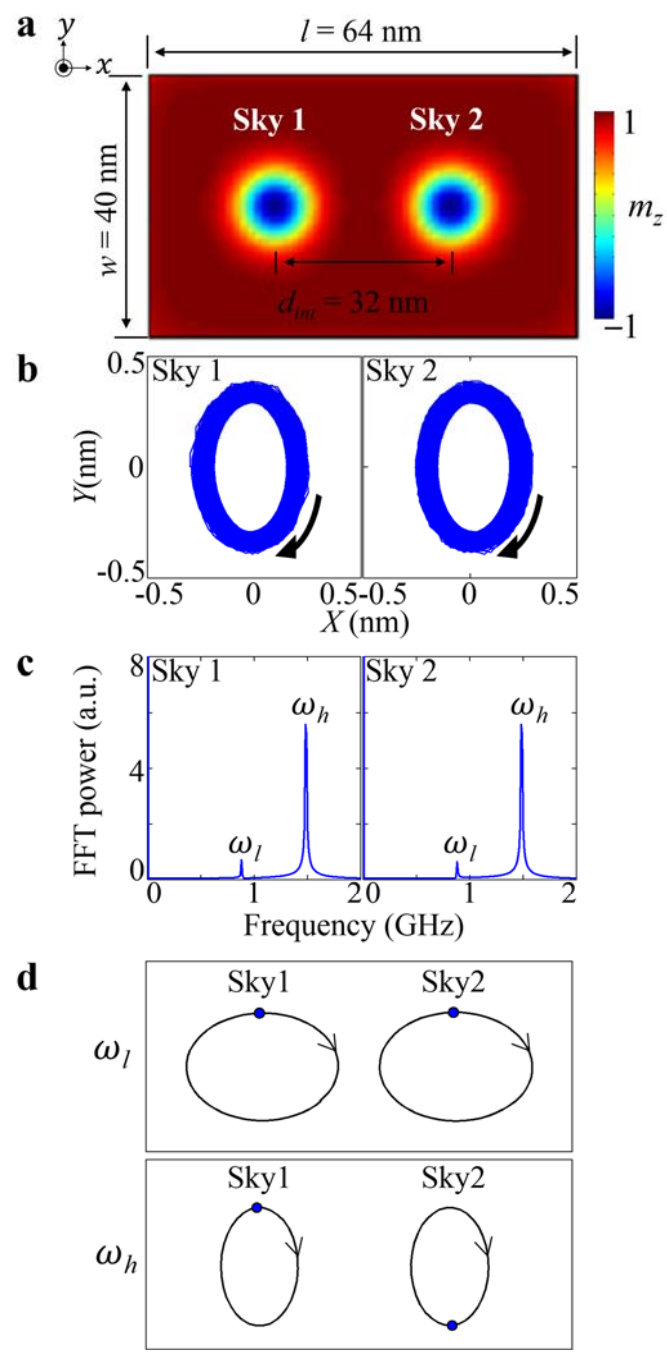

**Figure 3**

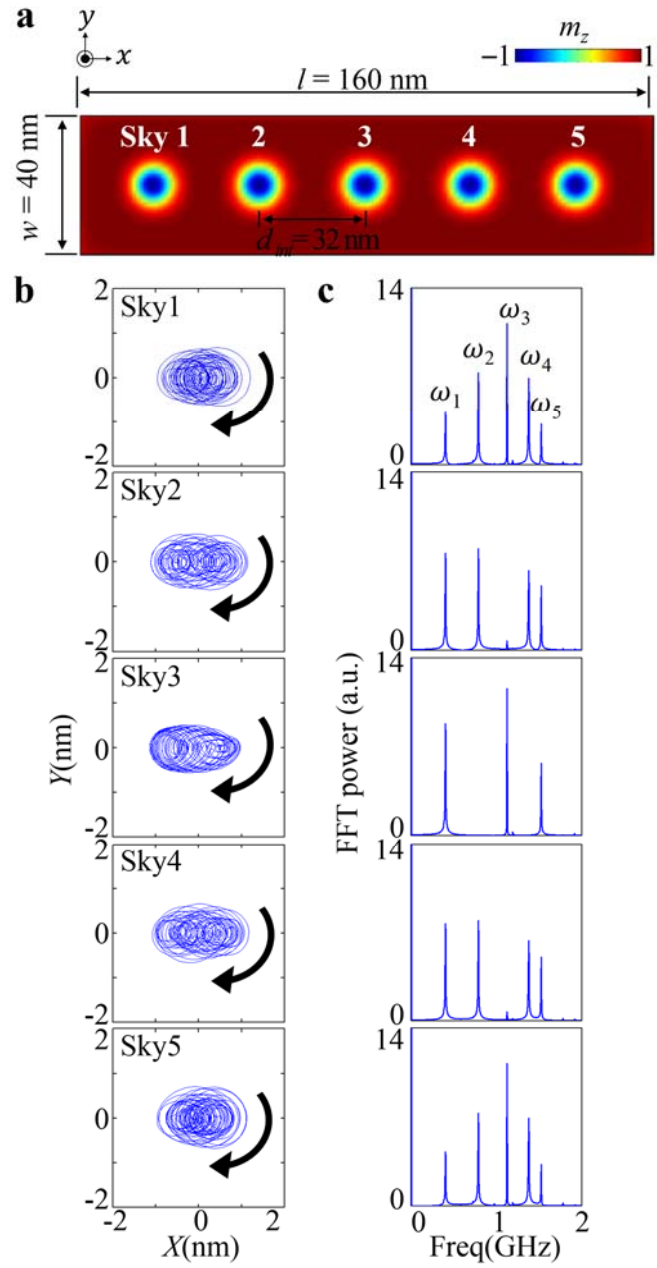

**Figure4**

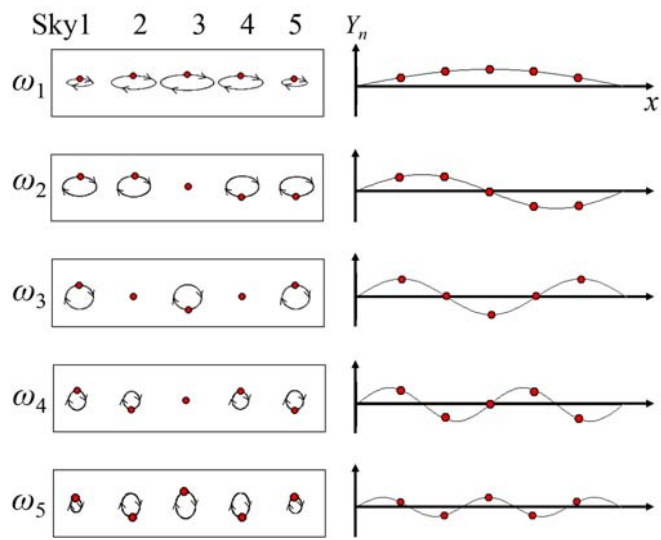

**Figure5**

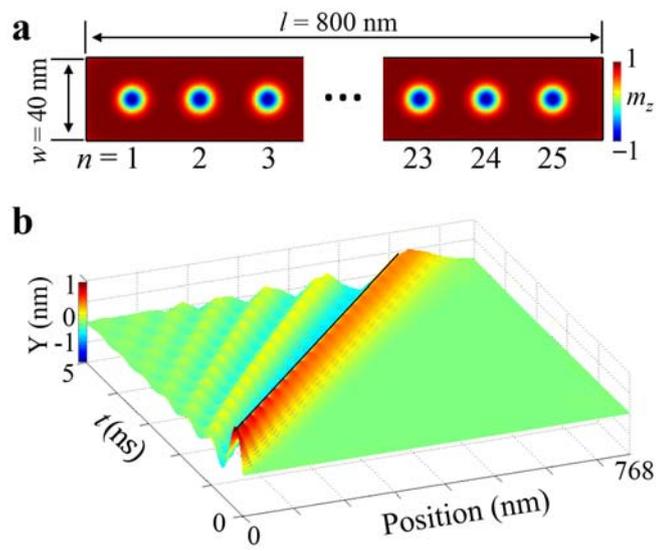

**Figure6**

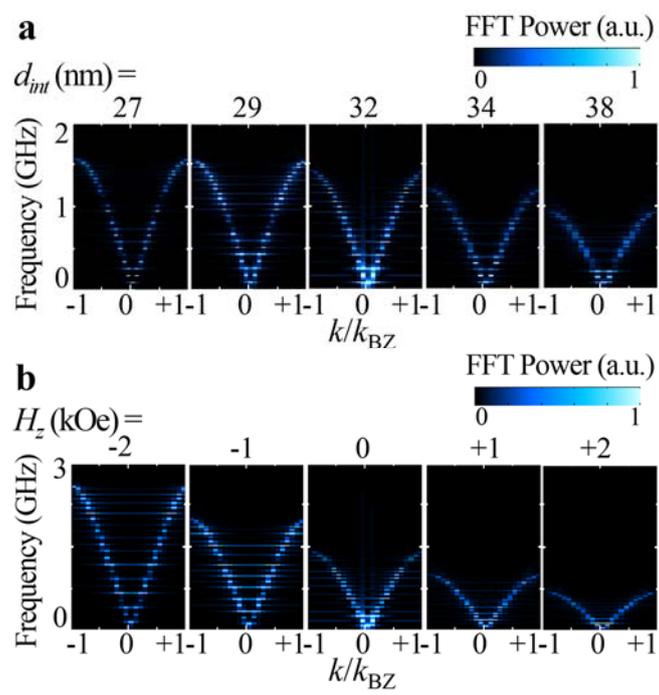

**Figure7**

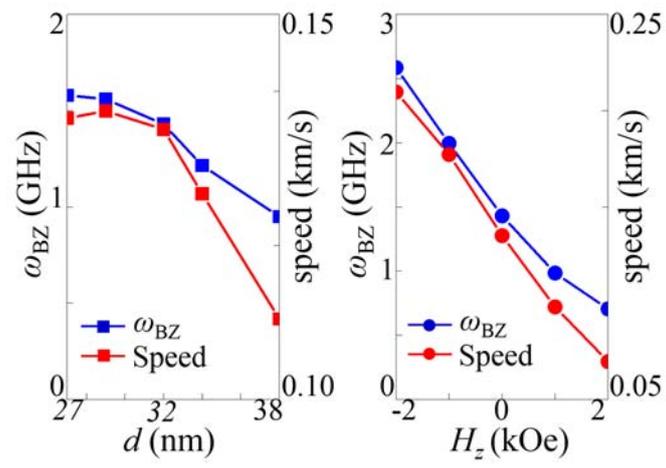

**SUPPLEMENTARY INFORMATION**

# Coupled gyration modes and propagation in one-dimensional periodic skyrmion arrays as reliable information carrier

Junhoe Kim, Jaehak Yang, Young-Jun Cho, Bosung Kim, and Sang-Koog Kim[a]

*National Creative Research Initiative Center for Spin Dynamics and Spin-Wave Devices, Nanospinics Laboratory, Research Institute of Advanced Materials, Department of Materials Science and Engineering, Seoul National University, Seoul 151-744, Republic of Korea*

Correspondence and requests for materials should be addressed to S.-K.Kim (sangkoog@snu.ac.kr).

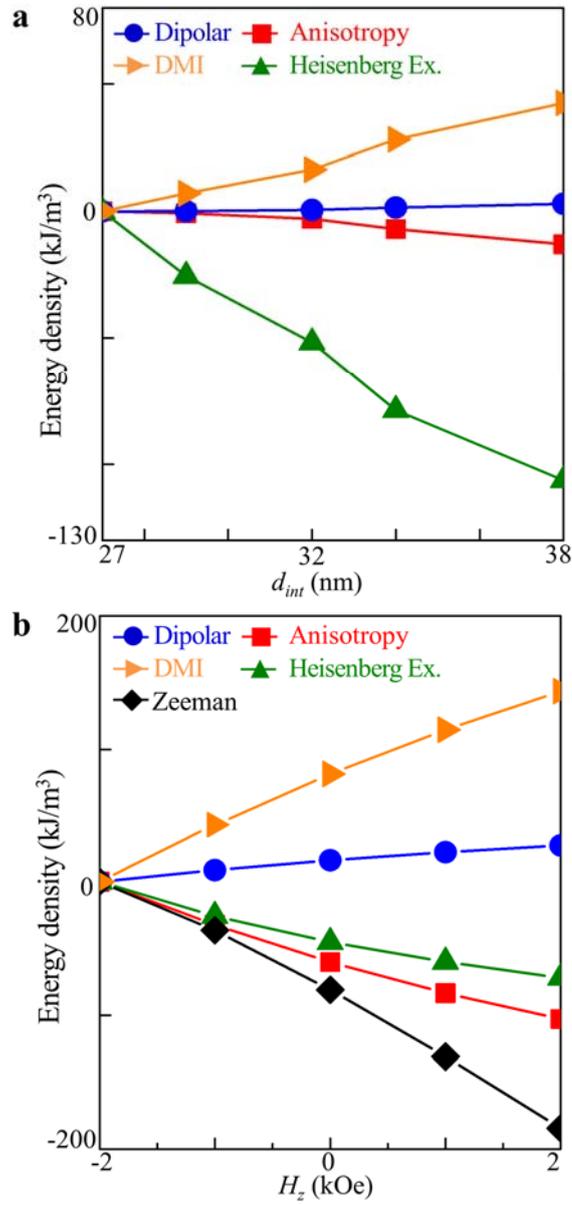

**Supplementary Fig. S1.** Individual terms of magnetic energy densities versus (a) $d_{int}$ in the given dimensions of nanostrip with $H_z$= 0 and versus (b) $H_z$ for the case of $d_{int}$= 32 nm

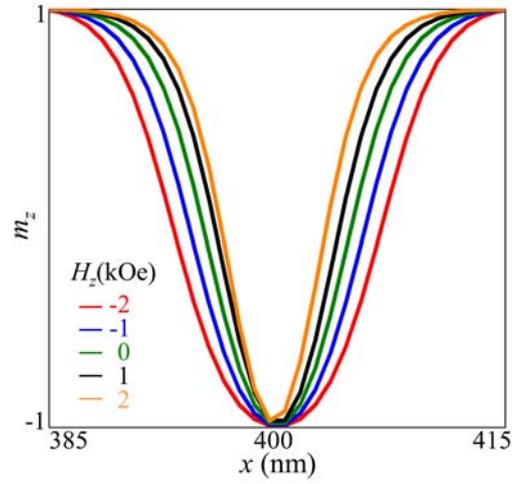

**Supplementary Fig. S2.** $m_z$ profiles of 13th skyrmion in center region of nanostrip ($x = 385$–415 nm) at $y = 20$ nm for $N = 25$ ($d_{int} = 32$ nm) for different values of perpendicular magnetic field, $H_z$

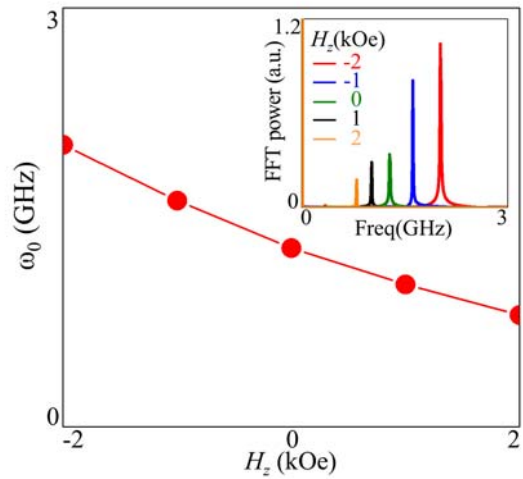

**Supplementary Fig. S3.** Eigenfrequency $\omega_0$ of CW gyration of single skyrmion in square dot versus $H_z$. The inset shows the FFT power spectra, as obtained from the FFTs of the temporal oscillations of the $y$ position of the skyrmion-core motion excited by the in-plane sinc field.

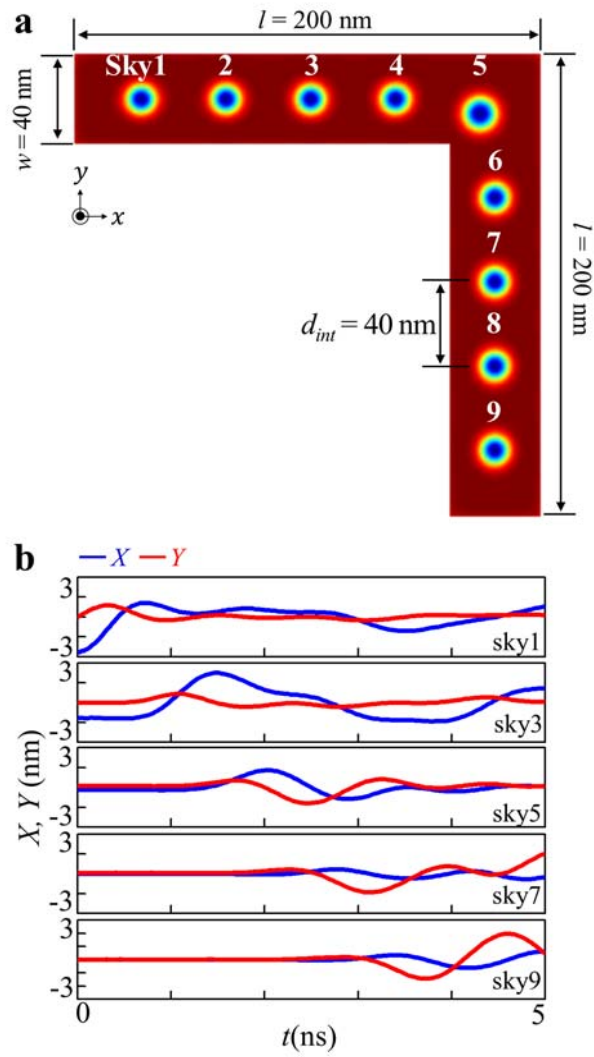

**Supplementary Fig. S4.** (a) Model geometry of 1D skyrmion array in an L-curved nanostrip. (b) Oscillatory $x$ (blue) and $y$ (red) components of core positions in the $n$th skyrmion.